\documentclass{llncs}
\usepackage{latexsym}

\newcommand{\almazero}{{\sf Alma-0}}

\newcommand{\ES}{\mbox{$\emptyset$}}

\newcommand{\ra}{\mbox{$\:\rightarrow\:$}}

\newcommand{\lra}{\mbox{$\:\leftrightarrow\:$}}

\newcommand{\A}{\mbox{$\ \wedge\ $}}

\newcommand{\Or}{\mbox{$\ \vee\ $}}

\newcommand{\sse}{\mbox{$\:\subseteq\:$}}

\newcommand{\fa}{\mbox{$\forall$}}
\newcommand{\te}{\mbox{$\exists$}}

\newcommand{\LL}{\mbox{$\ldots$}}

\newcommand{\newMS}[1]{\mbox{$[\![{#1}]\!]$}}
\newcommand{\newhat}[1]{\langle {#1} \rangle}

\newcommand{\B}[1]{\mbox{$[\![{#1}]\!]$}}       
\newcommand{\C}[1]{\mbox{$\{{#1}\}$}}           

\newcommand{\NI}{\noindent}
\newcommand{\HB}{\hfill{$\Box$}}
\newcommand{\VV}{\vspace{5 mm}}
\newcommand{\III}{\vspace{3 mm}}
\newcommand{\II}{\vspace{2 mm}}

\newcommand{\szkew}[1]{\relax \setbox0=\hbox{\kern -24pt $\displaystyle#1$\kern 0pt }%
\box0}
{\catcode`\@=11 \global\let\ifjusthvtest@=\iffalse}

\newcounter{oldmycaption}

\def\smallromani{\renewcommand{\theenumi}{\roman{enumi}}
\renewcommand{\labelenumi}{(\theenumi)}}

\newcommand{\Proof}{\NI
                    {\bf Proof.}\ }
\title{A Denotational Semantics for First-Order Logic}

\author{Krzysztof R. Apt\inst{1,2}}
\institute{CWI, P.O. Box 94079, 1090 GB Amsterdam, the Netherlands\\
\and 
University of Amsterdam, the Netherlands \\
http://www.cwi.nl/\~{}apt}

\begin{document}
\date{}
\maketitle

\begin{abstract}
In Apt and Bezem \cite{AB99} we provided a computational
interpretation of first-order formulas over arbitrary interpretations.
Here we complement this work by introducing a denotational semantics
for first-order logic.  Additionally, by allowing an assignment of a
non-ground term to a variable we introduce in this framework logical
variables.

The semantics combines a number of well-known ideas from the areas of
semantics of imperative programming languages and logic programming.
In the resulting computational view conjunction corresponds to
sequential composition, disjunction to ``don't know'' nondeterminism,
existential quantification to declaration of a local variable, and
negation to the ``negation as finite failure'' rule.  The soundness
result shows correctness of the semantics with respect to the notion
of truth.  The proof resembles in some aspects the proof of the
soundness of the SLDNF-resolution.

\end{abstract}

\section{Introduction}
\subsection*{Background}

To explain properly the motivation for the work here discussed we need
to go back to the roots of logic programming and constraint logic
programming.  {\em Logic programming\/} grew out of the seminal work
of Robinson \cite{Rob65} on the {\em resolution method\/} and the {\em
unification method}.  First, Kowalski and Kuehner \cite{KK71}
introduced a limited form of resolution, called linear
resolution. Then Kowalski \cite{Kow74} proposed what we now call {\em
SLD-resolution}.  The SLD-resolution is both a restriction and an
extension of the resolution method.  Namely, the clauses are
restricted to Horn clauses. However, in the course of the resolution
process a substitution is generated that can be viewed as a result of
a computation. Right from the outset the SLD-resolution became then a
crucial example of the {\em computation as deduction\/} paradigm
according to which the computation process is identified with a
constructive proof of a formula (a query) from a set of axioms (a
program) with the computation process yielding the witness (a
substitution).

This lineage of logic programming explains two of its relevant characteristics:

\begin{enumerate}

\item the queries and clause bodies are limited to the conjunctions of atoms,

\item the computation takes place (implicitly) over the domain of all ground terms
of a given first-order language.

\end{enumerate}

The restriction in item 1. was gradually lifted and through the works of
Clark \cite{Cla78} and Lloyd and Topor \cite{LT84}
one eventually arrived at the possibility of
using as queries and clause bodies arbitrary first-order formulas.
This general syntax is for example available in the language G\"{o}del of
Lloyd and Hill \cite{HL94}.

A way to overcome the restriction in item 2. was proposed in 1987 by
Jaffar and Lassez in their influential CLP(X) scheme that led to {\em
constraint logic programming}.  In this proposal the computation takes
place over an arbitrary interpretation and the queries and clause
bodies can contain constraints, i.e., atomic formulas interpreted over
the chosen interpretation.  The unification mechanism is replaced by a
more general process of constraint solving and the outcome of a
computation is a sequence of constraints to which the original query
reduces.

This powerful idea was embodied since then in many constraint logic
programming languages, starting with the CLP$({\cal R})$ language of
Jaffar, Michaylov, Stuckey, and Yap \cite{jaffar-clpr}
in which linear constraints over
reals were allowed, and the CHIP language of Dincbas et al.
\cite{dincbas88b}
in which linear constraints over finite domains, combined
with constraint propagation, were introduced.  A theoretical framework
for CHIP was provided in van Hentenryck \cite{vanhentenryck-constraint}.

This transition from logic programming to constraint logic programming
introduced a new element. In the  CLP(X) scheme the test for
satisfiability of a sequence of constraints was needed, while a proper
account of the CHIP computing process required an introduction of
constraint propagation into the framework.  On some interpretations these
procedures can be undecidable (the satisfiability test) or
computationally expensive (the ``ideal'' constraint propagation). This
explains why in the realized implementations some approximation of the
former or limited instances of the latter were chosen for.

So in both approaches the computation (i.e., the deduction) process
needs to be parametrized by external procedures that for each
specific interpretation have to be provided and implemented
separately.  In short, in both cases the computation process, while
parametrized by the considered interpretation, also depends on the
external procedures used.
In conclusion: constraint logic programming did not provide a
satisfactory answer to the question of how to lift the computation
process of logic programming from the domain of all ground terms to an
arbitrary interpretation without losing the property that this process
is effective.

Arbitrary interpretations are important since they represent a
declarative counterpart of data types.  In practical situations the
selected interpretations would admit sorts that would correspond to
the data types chosen by the user for the application at hand, say
terms, integers, reals and/or lists, each with the usual operations
available.
It is useful to contrast this view with the one taken in typed
versions of logic programming languages.  For example, in the case of
the G\"{o}del language (polymorphic) types are provided and are
modeled by (polymorphic) sorts in the underlying theoretic model.
However, in this model the computation still implicitly takes place
over one fixed domain, that of all ground terms partitioned into sorts.
This domain properly captures the built-in types but does not
provide an account of user defined types. Moreover, in this approach
different (i.e., not uniform) interpretation of equality for different
types is needed, a feature present in the language but not accounted
for in the theoretical model.

\subsection*{Formulas as Programs}

The above considerations motivated our work on a computational
interpretation of first-order formulas over arbitrary interpretations
reported in Apt and Bezem \cite{AB99}.  This allowed us to view
first-order formulas as executable programs.  That is why we called
this approach {\em formulas as programs}.  In our approach the
computation process is a search of a satisfying valuation
for the formula in question.  Because the problem of finding such a
valuation is in general undecidable, we had to introduce the
possibility of partial answers, modeled by an existence of run-time
errors.

This ability to compute over arbitrary interpretations allowed us to
extend the computation as deduction paradigm to arbitrary
interpretations.  We noted already that the SLD-resolution is both a
restriction and an extension of the resolution method.  In turn, the
formulas as programs approach is both a restriction and an extension
of the logic programming.  Namely, the unification process is limited
to an extremely simple form of matching involving variables and ground
terms only.  However, the computation process now takes place over an
arbitrary structure and full-first order syntax is adopted.

The formulas as programs approach to programming has been realized in
the programming language \almazero{} \cite{ABPS98a} that extends
imperative programming by features that support declarative
programming.  In fact, the work reported in Apt and Bezem \cite{AB99}
provided logical underpinnings for a fragment of \almazero{} that does
not include destructive assignment or recursive procedures and allowed
us to reason about non-trivial programs written in this fragment.

\subsection*{Rationale for This Paper}

The computational interpretation provided in Apt and Bezem \cite{AB99}
can be viewed as an operational semantics of first-order logic.  The
history of semantics of programming languages has taught us that to
better understand the underlying principles it is beneficial to
abstract from the details of the operational semantics.  This view was
put forward by Scott and Strachey \cite{SS71} in their proposal of {\em
denotational semantics\/} of programming languages according to which,
given a programming language, the meaning of each program is a
mathematical function of the meanings of its direct constituents.

The aim of this paper is to complement the work of \cite{AB99} by
providing a denotational semantics of first-order formulas.  This
semantics combines a number of ideas realized in the areas of
(nondeterministic) imperative programming languages and the field of
logic programming.  It formalizes a view according to which
conjunction can be seen as sequential composition, disjunction as
``don't know'' nondeterminism, existential quantification as
declaration of a local variable, and it relates negation to the
``negation as finite failure'' rule.

The main result is that the denotational semantics is sound with
respect to the truth definition. The proof is reminiscent in some
aspects of the proof of the soundness of the SLDNF-resolution of
Clarke \cite{Cla78}.  The semantics of equations allows matching
involving variables and non-ground terms, a feature not present in
\cite{AB99} and in \almazero{}. This facility introduces logical
variables in this framework but also creates a number of difficulties
in the soundness proof because bindings to local variables can now be
created.

First-order logic is obviously a too limited formalism for
programming. In \cite{AB99} we discussed a number of extensions
that are convenient for programming purposes, to wit sorts (i.e.,
types), arrays, bounded quantification and non-recursive procedures.
This leads to a very expressive and easy to program in subset of
\almazero{}. We do not envisage any problems in incorporating these
features into the denotational semantics here provided.  A major
problem is how to deal with recursion.

The plan of the paper is as follows. In the next section we discuss
the difficulties encountered when solving arbitrary equations over
algebras.  Then, in Section \ref{sec:equ} we provide a semantics of
equations and in Section \ref{sec:fol} we extend it to the case of
first-order formulas interpreted over an arbitrary interpretation.
The resulting semantics is denotational in style.  In Section
\ref{sec:soundness} we relate this semantics to the notion of truth by
establishing a soundness result.  In Section \ref{sec:conclusions} we
draw conclusions and suggest some directions for future work.

\section{Solving Equations over Algebras}

Consider some fixed, but arbitrary, {\em language of terms\/} $L$ and
a fixed, but arbitrary {\em algebra\/} ${\cal J}$ for it (sometimes called a
{\em pre-interpretation\/}).  A typical example is the language
defining arithmetic expressions and its standard interpretation over
the domain of integers.

We are interested in solving equations of the form $s = t$ over an
algebra, that is, we seek an instantiation of the variables occurring
in $s$ and $t$ that makes this equation true when interpreted over
${\cal J}$.  By varying $L$ and ${\cal J}$ we obtain a whole array of
specific decision problems that sometimes can be solved efficiently,
like the unification problem or the problem of solving linear
equations over reals, and sometimes are undecidable, like the problem
of solving Diophantine equations.

Our intention is to use equations as a means to assign values to
variables. Consequently, we wish to find a natural, general, situation
for which the problem of determining whether an equation $s = t$ has a
solution in a given algebra is decidable, and to exhibit a ``most general
solution'', if one exists.  By using most general solutions we do not
lose any specific solution.

This problem cannot be properly dealt with in full generality.  Take
for example the polynomial equations over integers.  Then the equation
$x^{2} - 3x + 2 = 0$ has two solutions, $\C{x/1}$ and $\C{x/2}$, and
none is ``more general'' than the other under any reasonable
definition of a solution being more general than another.


In fact, given an arbitrary interpretation, the only case that seems
to be of any use is that of comparing a variable and an arbitrary
term.  This brings us to equations of the form $x = t$, where $x$ does
not occur in $t$. Such an equation has obviously a most general
solution, namely the instantiation $\C{x/t}$.

A dual problem is that of finding when an equation $s = t$ has no
solution in a given algebra.  Of course, non-unifiability is not a
rescue here: just consider the already mentioned equation 
$x^{2} - 3x + 2 = 0$ the sides of which do not unify.

Again, the only realistic situation seems to be when both terms are
ground and their values in the considered algebra are different.  This
brings us to equations $s = t$ both sides of which are ground terms.

\section{Semantics of Equations}
\label{sec:equ}

After these preliminary considerations we introduce specific
``hybrid'' objects in which we mix the syntax and semantics.

\begin{definition} 
  Consider a language of terms $L$ and an algebra ${\cal J}$ for it.
  Given a function symbol $f$ we denote by $f_{\cal J}$ the 
  interpretation of $f$ in ${\cal J}$.

\begin{itemize}
\item Consider a term of $L$ in which we replace some of the variables
by the elements of the domain $D$. We call the resulting object
a {\em generalized term}.

\item Given a generalized term $t$ we define its {\em ${\cal J}$-evaluation\/} as follows:
\begin{itemize}

\item replace each constant occuring in $t$ by its value in ${\cal J}$,

\item repeatedly replace each sub-object of the form
$f(d_1, \LL, d_n)$ where $f$ is a function symbol and $d_1, \LL, d_n$ are the
elements of the domain $D$ by the element $f_{{\cal J}}(d_1, \LL, d_n)$ of $D$.
  \end{itemize}

We call the resulting generalized term a {\em ${\cal J}$-term} and denote it by $\B{t}_{\cal J}$.
Note that if $t$ is ground, then  $\B{t}_{\cal J}$ is an element of the domain of ${\cal J}$.

\item By a {\em ${\cal J}$-substitution\/} we mean a finite mapping
from variables to ${\cal J}$-terms which assigns to each variable $x$
in its domain a ${\cal J}$-term different from $x$.  We write it as
$\C{x_1/h_1,\dots,x_n/h_n}$.  
\HB
\end{itemize}
\end{definition}

The ${\cal J}$-substitutions generalize both the usual substitutions
and the valuations, which assign domain values to variables.
By adding to the language $L$ constants for each domain element
and for each ground term we can reduce the 
${\cal J}$-substitutions to the substitutions. We preferred not
to do this to keep the notation simple.

In what follows we denote the empty ${\cal J}$-substitution by
$\varepsilon$ and arbitrary ${\cal J}$-substitutions by $\theta, \eta,
\gamma$ with possible subscripts.

A more intuitive way of introducing ${\cal J}$-terms is as follows.  Each
ground term of $s$ of $L$ evaluates to a unique value in ${\cal J}$. Given a
generalized term $t$ replace each maximal ground subterm of $t$ by its
value in ${\cal J}$.  The outcome is the ${\cal J}$-term $\B{t}_{\cal J}$.

We define the notion of an application of a ${\cal J}$-substitution
$\theta$ to a generalized term $t$ in the standard way and denote it
by $t \theta$.  If $t$ is a term, then $t\theta$ does not have to be a
term, though it is a generalized term.

\VV

\begin{definition} \mbox{} \\
\vspace{-6mm}

\begin{itemize}


\item A {\em composition of two ${\cal J}$-substitutions $\theta$
and $\eta$}, written as $\theta \eta$, is defined as the unique
${\cal J}$-substitution $\gamma$ such that for each variable $x$
\[
x \gamma = \B{(x \theta) \eta}_{\cal J}. 
\]
\HB
\end{itemize}
\end{definition}

Let us illustrate the introduced concepts by means of two examples.

\begin{example} \label{exa:1}
Take an arbitrary language of terms $L$. The {\em Herbrand algebra} $Her$
for $L$ is defined as follows:
\begin{itemize}
\item
its domain is the set ${HU}_{L}$ of all ground terms of $L$ 
(usually called the {\em Herbrand universe}),

\item if $f$ is an $n$-ary function symbol in $L$, then its
interpretation is the mapping from $({HU}_{L})^n$ to ${HU}_{L}$ which
maps the sequence $t_1 ,\dots,t_n$ of ground terms to the ground term
$f(t_1,\dots,t_n)$.
\end{itemize}

Consider now a term $s$.  Then $\B{s}_{Her}$ equals $s$ because in
$Her$ every ground term evaluates to itself.
So the notions of a term, a generalized term and a $Her$-term coincide.
Consequently, the notions of substitutions and $Her$-substitutions
coincide.
\HB
\end{example}

\begin{example} \label{exa:2}
  Take as the language of terms the language $AE$ of {\em arithmetic
    expressions}.  Its binary function symbols are the usual $\cdot$
  (``times''), $+$ (``plus'') and $-$ (``minus''), and its unique binary
  symbol is $-$ (``unary minus'').  Further, for each integer ${\bf
    k}$ there is a constant $k$.

As the algebra for $AE$ we choose the standard algebra $Int$ that
consists of the set of integers with the function symbols interpreted
in the standard way.  In what follows we write the binary function
symbols in the usual infix notation.

Consider the term $s \equiv x + (((3 + 2) \cdot 4) - y)$.  Then
$\B{s}_{AE}$ equals $x + ({\bf 20} - y)$.  Further, given the
$AE$-substitution $\theta := \C{x/{\bf 6}-z, \: y/{\bf 3}}$ we have
$s\theta \equiv ({\bf 6}-z) + (((3 + 2) \cdot 4) - {\bf 3})$ and
consequently, $\B{s\theta}_{AE} = ({\bf 6}-z) + {\bf 17}$.
Further, given $\eta := \C{z/{\bf 4}}$, we have
$\theta \eta = \C{x/{\bf 2}, \: y/{\bf 3}, \: z/{\bf 4}}$. 
\HB
\end{example}

To define the meaning of an equation over an algebra ${\cal J}$
we view ${\cal J}$-substi\-tu\-tions as states and use a special state
\begin{itemize}

\item {\em error}, to indicate that it is not possible to determine
effectively
  whether a solution to the equation $s\theta =t\theta$ in ${\cal J}$ exists.
\end{itemize}

We now define the semantics $\newMS{\cdot}$ of an equation between
two generalized terms as follows:

\begin{eqnarray*}
\begin{array}{lll}
\newMS{s=t}(\theta) & :=  &
\left\{
\begin{array}{ll}
\C{\theta \C{s\theta/\B{t\theta}_{\cal J}}} &  \mbox{if $s\theta$ is a variable 
that does not occur in $t\theta$,} \\
\C{\theta \C{t\theta/\B{s\theta}_{\cal J}}} &  \mbox{if $t\theta$ is a variable 
that does not occur in $s\theta$} \\
                                           &   \mbox{and $s\theta$ is not a variable,} \\
\C{\theta} &  \mbox{if $\B{s\theta}_{\cal J}$ and $\B{t\theta}_{\cal J}$ are identical,} \\
\ES &  \mbox{if $s\theta$ and $t\theta$ are ground and $\B{s\theta}_{\cal J} \neq \B{t\theta}_{\cal J}$,} \\
\C{error} &  \mbox{otherwise.}
\end{array}
\right.
\end{array}
\end{eqnarray*}

It will become clear in the next section why we collect here the unique
outcome into a set and why we ``carry'' $\theta$ in the answers.

Note that according to the above definition we have
$\newMS{s=t}(\theta) = \C{error}$ for the non-ground generalized terms
$s\theta$ and $t\theta$ such that the ${\cal J}$-terms
$\B{s\theta}_{\cal J}$ and $\B{t\theta}_{\cal J}$ are different.  In
some situations we could safely assert then that $\newMS{s=t}(\theta)
= \C{\theta}$ or that $\newMS{s=t}(\theta) = \ES$.  For example, for
the standard algebra $Int$ for the language of arithmetic expressions
we could safely assert that $\newMS{x+x = 2 \cdot x}(\theta) =
\C{\theta}$ and $\newMS{x+1 = x}(\theta) = \ES$ for any
$AE$-substitution $\theta$.

The reason we did not do this was that we wanted to ensure that the
semantics is uniform and decidable so that it can be implemented.

\section{A Denotational Semantics for First-Order Logic}
\label{sec:fol}

Consider now a first-order language with equality $\cal L$.  In this
section we extend the semantics $\newMS{\cdot}$ to arbitrary
first-order formulas from $\cal L$ interpreted over an arbitrary
interpretation.  $\newMS{\cdot}$ depends on the considered
interpretation but to keep the notation simple we do not indicate this
dependence. This semantics is denotational in the sense that meaning
of each formula is a mathematical function of the meanings of its
direct constituents.

Fix an interpretation ${\cal I}$. ${\cal I}$ is based on some algebra
${\cal J}$.  We define the notion of an application of a ${\cal
  J}$-substitution $\theta$ to a formula $\phi$ of $\cal L$, written
as $\phi \theta$, in the usual way.

Consider an atomic formula $p(t_1, \LL, t_n)$ and a ${\cal
  J}$-substitution $\theta$.  We denote by $p_{\cal I}$ the
interpretation of $p$ in ${\cal I}$.

We say that

\begin{itemize}

\item $p(t_1, \LL, t_n)\theta$ is {\em true\/} if
$p(t_1, \LL, t_n)\theta$ is ground
and $(\B{t_{1}\theta}_{\cal J}, \LL, \B{t_{n}\theta}_{\cal J}) \in p_{\cal I}$,

\item $p(t_1, \LL, t_n)\theta$ is {\em false\/} if
$p(t_1, \LL, t_n)\theta$ is ground
and $(\B{t_{1}\theta}_{\cal J}, \LL, \B{t_{n}\theta}_{\cal J}) \not\in p_{\cal I}$.

\end{itemize}

In what follows we denote by $Subs$ the set of ${\cal J}$-substitutions
and by ${\cal P}(A)$, for a set $A$, the set of all subsets of $A$.

For a given formula $\phi$ its semantics $\newMS{\phi}$ is a mapping
\[
\newMS{\phi} : Subs \ra {\cal P}(Subs \cup \C{error}).
\]

The fact that the outcome of $\newMS{\phi}(\theta)$ is a set reflects
the possibility of a nondeterminism here modeled by the disjunction.

To simplify the definition we extend $\newMS{\cdot}$ to deal with subsets
of $Subs \cup \C{error}$ by putting 
\[
\newMS{\phi}(error) := \C{error},
\]
and for a set $X \sse Subs \cup \C{error}$
\[
\newMS{\phi}(X) := \bigcup_{e \in X} \newMS{\phi}(e).
\]

Further, to deal with the existential quantifier, we introduce an
operation $DROP_x$, where $x$ is a variable.
First we define $DROP_x$ on the elements of 
$Subs \cup \C{error}$ by putting for a ${\cal J}$-substitution
$\theta$
\begin{eqnarray*}
\begin{array}{lll}
DROP_{x}(\theta) & :=  &
\left\{
\begin{array}{ll}
\theta  &  \mbox{if $x$ is not in the domain of $\theta$,} \\
\eta  &  \mbox{if $\theta$ is of the form $\eta \uplus \C{x/s}$,}
\end{array}
\right.
\end{array}
\end{eqnarray*}
and
\[
DROP_{x}(error) := error.
\]

Then we extend it element-wise to subsets of $Subs \cup \C{error}$,
that is, by putting for a set $X \sse Subs \cup \C{error}$
\[
DROP_{x}(X) := \C{DROP_{x}(e) \mid e \in X}.
\]

$\newMS{\cdot}$ is defined by structural induction as follows,
where $A$ is an atomic formula different from $s=t$:

\begin{itemize}

\item $\newMS{A}(\theta) :=
\left\{
\begin{array}{ll}
\C{\theta} &  \mbox{if $A \theta$ is true,} \\
\ES &  \mbox{if $A \theta$ is false,} \\
\C{error} &  \mbox{otherwise, that is if $A \theta$ is not ground,}
\end{array}
\right.
$

\item $\newMS{\phi_1 \A \phi_2}(\theta) :=\newMS{\phi_2}(\newMS{\phi_1}(\theta))$,

\item $\newMS{\phi_1 \Or \phi_2}(\theta) :=\newMS{\phi_1}(\theta) \cup \newMS{\phi_2}(\theta)$,

\item $\newMS{\neg \phi}(\theta) :=
\left\{
\begin{array}{ll}
\C{\theta} &  \mbox{if $\newMS{\phi}(\theta) = \ES$,} \\
\ES &  \mbox{if $\theta \in \newMS{\phi}(\theta)$,} \\
\C{error} &  \mbox{otherwise,}
\end{array}
\right.
$
\item $\newMS{\te x \: \phi}(\theta) := DROP_{y}(\newMS{\phi\C{x/y}}(\theta))$,
where $y$ is a fresh variable.

\end{itemize}

To better understand this definition let us consider some simple
examples that refer to the algebras discussed in 
Examples \ref{exa:1} and \ref{exa:2}.

\begin{example} \label{exa:1a}

Take an interpretation ${\cal I}$ based on the Herbrand
algebra $Her$. Then
\[
\newMS{f(x) = z \A g(z) = g(f(x))}(\C{x/g(y)}) =
\newMS{g(z) = g(f(x))}(\theta) = \C{\theta},
\]
where $\theta := \C{x/g(y), z/f(g(y))}$.
On the other hand
\[
\newMS{g(f(x)) = g(z)}(\C{x/g(y)}) = \C{error}.
\]
\HB
\end{example}

\begin{example} \label{exa:2a}

Take an interpretation ${\cal I}$ based on the standard
algebra $AE$ for the language of arithmetic expressions. Then
\[
\newMS{y = z-1 \A z = x+2}(\C{x/{\bf 1}}) =
\newMS{z = x+2}(\C{x/{\bf 1}, y/z-{\bf 1}}) =
\C{x/{\bf 1}, y/{\bf 2}, z/{\bf 3}}.
\]
Further,
\[
\newMS{y+1 = z-1}(\C{y/{\bf 1}, z/{\bf 3}}) =
\C{y/{\bf 1}, z/{\bf 3}}
\]
and even
\[
\newMS{x \cdot (y+1) = (v+1) \cdot (z-1)}(\C{x/v + {\bf 1}, \: y/{\bf 1}, \: z/{\bf 3}}) =
\C{x/v + {\bf 1}, \: y/{\bf 1}, \: z/{\bf 3}}.
\]
On the other hand
\[
\newMS{y-1 = z-1}(\varepsilon) = \C{error}.
\]
\HB
\end{example}

The first example shows that the semantics given here is weaker
than the one provided by the logic programming.  In turn, the second
example shows that our treatment of arithmetic expressions is more
general than the one provided by Prolog.

This definition of denotational semantics of first-order formulas
combines a number of ideas put forward in the area of semantics of
imperative programming languages and the field of logic programming.

First, for an atomic formula $A$, when $A\theta$ is ground, its
meaning coincides with the meaning of a Boolean expression
given in de Bakker \cite[page 270]{Bak80}.
In turn, the meaning of the conjunction and of the disjunction follows
\cite[page 270]{Bak80} in the sense that the conjunction
corresponds to the sequential composition operation ``;'' and the
disjunction corresponds to the ``don't know'' nondeterministic choice,
denoted there by $\cup$.

Next, the meaning of the negation is inspired by its treatment in logic
programming. To be more precise we need the following observations the
proofs of which easily follow by structural induction.

\begin{note} \label{not:extends} \mbox{} \\
\vspace{-6mm}

\begin{enumerate}\smallromani
\item If $\eta \in \newMS{\phi}(\theta)$, then $\eta = \theta \gamma$
for some ${\cal J}$-substitution $\gamma$.

\item If $\phi \theta$
is ground, then  $\newMS{\phi}(\theta) \sse \C{\theta}$.
\HB
\end{enumerate}
\end{note}

First, we interpret $\newMS{\phi}(\theta) \cap Subs \neq \ES$ as the
statement ``the query $\phi\theta$ succeeds''.  More specifically, if
$\eta \in \newMS{\phi}(\theta)$, then by Note \ref{not:extends}(i) for
some $\gamma$ we have $\eta = \theta \gamma$.

In general, $\gamma$ is of course not unique: take for example $\theta
:= \C{x/0}$ and $\eta = \theta$. Then both $\eta = \theta \varepsilon$
and $\eta = \theta \theta$. However, it is easy to show that if $\eta$
is less general than $\theta$, then in the set $\C{\gamma \mid \eta =
\theta \gamma}$ the ${\cal J}$-substitution with the smallest domain
is uniquely defined.  In what follows given ${\cal J}$-substitutions
$\eta$ and $\theta$ such that $\eta$ is less general than $\theta$,
when writing $\eta = \theta \gamma$ we always refer to this uniquely
defined $\gamma$.

Now we interpret $\theta \gamma \in \newMS{\phi}(\theta)$ as the
statement ``$\gamma$ is the computed answer substitution for the query
$\phi\theta$''.  In turn, we interpret $\newMS{\phi}(\theta) = \ES$ as
the statement ``the query $\phi\theta$ finitely fails''.

Suppose now that $\newMS{\phi}(\theta) \cap Subs \neq \ES$, which means
that the query $\phi\theta$ succeeds.  Assume additionally that $\phi
\theta$ is ground. Then by Note \ref{not:extends}(ii)
$\theta \in \newMS{\phi}(\theta)$ and consequently by the definition of
the meaning of negation $\newMS{\neg \phi}(\theta) = \ES$, which
means that the query  $\neg \phi\theta$ finitely fails.  

In turn, suppose that $\newMS{\phi}(\theta) = \ES$, which means that
the query $\phi\theta$ finitely fails. By the definition of the
meaning of negation $\newMS{\neg \phi}(\theta) = \C{\theta}$, which means
that the query $\neg \phi\theta$ succeeds with the empty computed
answer substitution.

This explains the relation with
the ``negation as finite failure'' rule according to which for a ground query $Q$:
\begin{itemize}

\item if $Q$ succeeds, then $\neg Q$ finitely fails,

\item if $Q$ finitely fails, then $\neg Q$ succeeds with the empty
computed answer substitution.
\end{itemize}

In fact, our definition of the meaning of negation corresponds to a
generalization of the negation as finite failure rule already mentioned in
Clark \cite{Cla78}, according to which the requirement that $Q$ is
ground is dropped and the first item is replaced by:

\begin{itemize}

\item if $Q$ succeeds with the empty computed answer substitution,
then $\neg Q$ finitely fails.
\end{itemize}

Finally, the meaning of the existential quantification corresponds to
the meaning of the block statement in imperative languages, see, e.g.,
de Bakker \cite[page 226]{Bak80}, with the important difference that the local
variable is not initialized.  From this viewpoint the existential
quantifier $\te x$ corresponds to the declaration of the local
variable $x$.  The $DROP_x$ operation was introduced in Clarke
\cite{Cla79} to deal with the declarations of local variables.

We do not want to make
the meaning of the formula $\te x \: \phi$ dependent
on the choice of $y$. Therefore we postulate that for {\em any\/}
fresh variable $y$ the set $DROP_{y}(\newMS{\phi\C{x/y}}(\theta))$ is
a meaning of $\te x \: \phi$ given a ${\cal J}$-substitution
$\theta$.  Consequently, the semantics of $\te x \: \phi$ has many
outcomes, one for each choice of $y$. This ``multiplicity'' of
meanings then extends to all formulas containing the existential
quantifier.  So for example for any variable $y$ different from $x$
and $z$ the ${\cal J}$-substitution $\C{z/f(y)}$ is the meaning of
$\te x \: (z = f(x))$ given the empty ${\cal J}$-substitution
$\varepsilon$.

\section{Soundness}
\label{sec:soundness}

To relate the introduced semantics to the notion of truth we first
formalize the latter using the notion of a ${\cal J}$-substitution
instead of the customary notion of a valuation.

Consider a first-order language ${\cal L}$ with equality and an
interpretation ${\cal I}$ for it based on some algebra ${\cal J}$.
Let $\theta$ be a ${\cal J}$-substitution.  We define the relation ${\cal
  I} \models_{\theta} \phi$ for a formula $\phi$ by structural
induction. First we assume that $\theta$ is defined on all free variables
of $\phi$ and put

\begin{itemize}
\item  ${\cal I} \models_{\theta} s = t$ iff $\B{s\theta}_{\cal J}$ and
  $\B{t\theta}_{\cal J}$ coincide,

\item  ${\cal I} \models_{\theta} p(t_1, \LL, t_n)$ iff 
$p(t_1, \LL, t_n)\theta$ is ground and $(\B{t_{1}\theta}_{\cal J}, \LL, \B{t_{n}\theta}_{\cal J}) \in p_{\cal I}$.
\end{itemize}
In other words, ${\cal I} \models_{\theta} p(t_1, \LL, t_n)$ iff
$p(t_1, \LL, t_n)\theta$ is true.  The definition extends to
non-atomic formulas in the standard way.

Now assume that $\theta$ is not defined on all free variables of $\phi$.
We put

\begin{itemize}
\item  ${\cal I} \models_{\theta} \phi$ iff ${\cal I} \models_{\theta} \fa x_1, \LL, \fa x_n \phi$
where $x_1, \LL, x_n$ is the list of the free variables of $\phi$ that do not occur in the domain
of $\theta$.
\end{itemize}

Finally, 
\begin{itemize}
\item  ${\cal I} \models \phi$ iff ${\cal I} \models_{\theta} \phi$ for all
${\cal J}$-substitutions $\theta$.
\end{itemize}

To prove the main theorem we need the following notation.
Given a ${\cal J}$-substitution $\eta := \C{x_1/h_1,\dots,x_n/h_n}$ we 
define $\newhat{\eta} := x_1 = h_1 \A \dots \A x_n = h_n$.

In the discussion that follows the following simple observation will
be useful.
\begin{note} \label{not:equality}
For all ${\cal J}$-substitutions $\theta$ and formulas $\phi$
\[
{\cal I} \models_{\theta} \phi \mbox{ iff } {\cal I} \models \newhat{\theta} \ra \phi.
\]
\HB
\end{note}

The following theorem now shows correctness of the introduced semantics 
with respect to the notion of truth.

\begin{theorem}[Soundness]\label{thm:soundness}

Consider a first-order language ${\cal L}$ with equality and an
interpretation ${\cal I}$ for it based on some algebra ${\cal J}$.
Let $\phi$ be a formula of  ${\cal L}$ 
and $\theta$ a ${\cal J}$-substitution.

\begin{enumerate}\smallromani
\item 
For each ${\cal J}$-substitution $\eta \in \newMS{\phi}(\theta)$
\[
{\cal I} \models_{\eta}\phi.
\]

\item 
If $error \not\in \newMS{\phi}(\theta)$, 
then 
\[
{\cal I} \models \phi \theta \lra \bigvee_{i = 1}^{k} \te {\bf y}_i \newhat{\eta_i},
\]
where $\newMS{\phi}(\theta) = \C{\theta \eta_1, \LL, \theta \eta_k}$,
and for $i \in [1..k]$ $\ {\bf y}_i$ is a sequence of variables that
appear in the range of $\eta_i$.

\end{enumerate}
\end{theorem}
Note that by $(ii)$ if
$\newMS{\phi}(\theta) = \ES$, then

\[
{\cal I} \models_{\theta} \neg \phi.
\]
In particular, if $\newMS{\phi}(\varepsilon) = \ES$, then
\[
{\cal I} \models \neg \phi.
\]
\Proof 
The proof proceeds by simultaneous induction on the structure of the formulas.
\III

\NI
$\phi$ is $s = t$.

If $\eta \in \newMS{\phi}(\theta)$, then three possibilities arise.
\II

\NI
1. $s\theta$ is a variable that does not occur in $t\theta$. 

Then 
$\newMS{s=t}(\theta) = \C{\theta \C{s\theta/\B{t\theta}_{\cal J}}}$ 
and consequently $\eta = \theta \C{s\theta/\B{t\theta}_{\cal J}}$.  
So ${\cal I} \models_{\eta} (s =
t)$ holds since $s \eta = \B{t\theta}_{\cal J}$ and $t \eta =
t\theta$.  
\II

\NI
2. $t\theta$ is a variable 
that does not occur in $s\theta$ and $s\theta$ is not a variable.

Then $\newMS{s=t}(\theta) = \C{\theta \C{t\theta/\B{s\theta}_{\cal J}}}$.
This case is symmetric to 1.
\II

\NI
3. $\B{s\theta}_{\cal J}$ and $\B{t\theta}_{\cal J}$ are identical.

Then $\eta = \theta$, so ${\cal I} \models_{\eta} (s = t)$ holds.
\III

If $error \not\in \newMS{\phi}(\theta)$, then four possibilities arise.
\II

\NI
1. $s\theta$ is a variable that does not occur in $t\theta$. 

Then 
$\newMS{s=t}(\theta) = \C{\theta \C{s\theta/\B{t\theta}_{\cal J}}}$.
We have 
${\cal I} \models (s = t) \theta \lra s\theta = \B{t\theta}_{\cal J}$.
\II

\NI
2. $t\theta$ is a variable 
that does not occur in $s\theta$ and $s\theta$ is not a variable.

Then $\newMS{s=t}(\theta) = \C{\theta \C{t\theta/\B{s\theta}_{\cal J}}}$.
This case is symmetric to 1.
\II

\NI
3. $\B{s\theta}_{\cal J}$ and $\B{t\theta}_{\cal J}$ are identical.

Then $\newMS{s=t}(\theta) = \C{\theta}$. We have $\newMS{s=t}(\theta) =
\C{\theta \varepsilon}$ and ${\cal I} \models_{\theta} s=t$, so
${\cal I} \models (s=t) \theta \lra \newhat{\varepsilon}$, since
$\newhat{\varepsilon}$ is vacuously true.  
\II

\NI
4. $s\theta$ and $t\theta$ are ground ${\cal J}$-terms and $\B{s\theta}_{\cal J} \neq \B{t\theta}_{\cal J}$.

Then $\newMS{s=t}(\theta) = \ES$ and
${\cal I} \models_{\theta} \neg (s=t)$, so
${\cal I} \models (s=t) \theta \lra falsum$,
where $falsum$ denotes the empty disjunction.
\III

\NI
$\phi$ is an atomic formula different from $s=t$.

If $\eta \in \newMS{\phi}(\theta)$, then 
$\eta = \theta$ and $\phi \theta$ is true. So ${\cal I} \models_{\theta} \phi$,
i.e., ${\cal I} \models_{\eta} \phi$.

If $error \not\in \newMS{\phi}(\theta)$, 
then either $\newMS{\phi}(\theta) = \C{\theta}$ or $\newMS{\phi}(\theta) = \ES$.
In both cases the argument is the same as in case 3. and 4. 
for the equality $s = t$.
\III

Note that in both cases we established a stronger form of $(ii)$ in
which each list ${\bf y}_i$ is empty, i.e., no quantification
over the variables in ${\bf y}_i$ appears.  
\III

\NI
$\phi$ is $\phi_1 \A \phi_2$. This is the most elaborate case.

If $\eta \in \newMS{\phi}(\theta)$, then for some ${\cal
J}$-substitution $\gamma$ both $\gamma \in \newMS{\phi_1}(\theta)$ and
$\eta \in \newMS{\phi_2}(\gamma)$.  By induction hypothesis both
${\cal I} \models_{\gamma}\phi_1$ and ${\cal I} \models_{\eta}\phi_2$.
But by Note \ref{not:extends}(i) $\eta$ is less general than $\gamma$,
so ${\cal I} \models_{\eta}\phi_1$ and consequently ${\cal I}
\models_{\eta}\phi_1 \A \phi_2$.

If $error \not\in  \newMS{\phi}(\theta)$, then for some $X \sse Subs$
both $\newMS{\phi_1}(\theta) = X$ and 
$error \not\in \newMS{\phi_2}(\eta)$ for all $\eta \in X$.

By induction hypothesis
\[
{\cal I} \models \phi_1 \theta \lra 
\bigvee_{i = 1}^{k} \te {\bf y}_i \newhat{\eta_i},
\]
where $X = \C{\theta \eta_1, \LL, \theta \eta_k}$
and for $i \in [1..k]$  $\ {\bf y}_i$ is a sequence of variables that
appear in the range of $\eta_i$.
Hence
\[
{\cal I} \models (\phi_1 \A \phi_2) \theta \lra 
\bigvee_{i = 1}^{k} (\te {\bf y}_i \newhat{\eta_i} \A \phi_2 \theta),
\]
so by appropriate renaming of the variables in the sequences ${\bf y}_i$
\[
{\cal I} \models (\phi_1 \A \phi_2) \theta \lra 
\bigvee_{i = 1}^{k} \te {\bf y}_i (\newhat{\eta_i} \A \phi_2 \theta).
\]

But for any ${\cal J}$-substitution $\delta$ and a formula $\psi$
\[
{\cal I} \models \newhat{\delta} \A \psi \lra \newhat{\delta} \A \psi \delta,
\]
so
\begin{equation}
{\cal I} \models (\phi_1 \A \phi_2) \theta \lra (
\bigvee_{i = 1}^{k} 
\te {\bf y}_i (\newhat{\eta_i} \A \phi_2 \theta \eta_i).
\label{eq:phi1}
\end{equation}

Further, we have for $i \in [1..k]$ 
\[
\newMS{\phi_2}(\theta \eta_i) = \C{\theta \eta_i \gamma_{i,j} \mid j \in [1 .. \ell_i]}
\]
for some ${\cal J}$-substitutions  $\gamma_{i,1}, \LL, \gamma_{i, \ell_i}$.
So
\[
\newMS{\phi_1 \A \phi_2}(\theta) = \C{\theta \eta_i \gamma_{i,j} \mid i \in [1..k], j \in [1 .. \ell_i]}.
\]
By induction hypothesis we have for $i \in [1..k]$ 
\[
{\cal I} \models \phi_2 \theta \eta_i \lra \bigvee_{j = 1}^{\ell_i} \te {\bf v}_{i,j}
\newhat{\gamma_{i,j}},
\]
where for $i \in [1..k]$ and $j \in [1.. \ell_i]$ $\ {\bf v}_{i,j}$ is
a sequence of variables that appear in the range of $\gamma_{i,j}$.

Using (\ref{eq:phi1}) by appropriate renaming of the variables in the sequences ${\bf v}_{i,j}$
we now conclude that
\[
{\cal I} \models (\phi_1 \A \phi_2) \theta \lra 
\bigvee_{i = 1}^{k}
\bigvee_{j = 1}^{\ell_i}
\te {\bf y}_{i} 
\te {\bf v}_{i,j}
(\newhat{\eta_i} \A \newhat{\gamma_{i,j}}),
\]
so
\[
{\cal I} \models (\phi_1 \A \phi_2) \theta \lra 
\bigvee_{i = 1}^{k}
\bigvee_{j = 1}^{\ell_i}
\te {\bf y}_{i} 
\te {\bf v}_{i,j}
\newhat{\eta_i \gamma_{i,j}},
\]
since the domains of $\eta_i$ and $\gamma_{i,j}$ are disjoint and for 
any ${\cal J}$-substitutions  $\gamma$ and $\delta$ with disjoint domains
we have
\[
{\cal I} \models \newhat{\gamma} \A \newhat{\delta} \lra \newhat{\gamma \delta}.
\]

\III

\NI
$\phi$ is $\phi_1 \Or \phi_2$.

If $\eta \in \newMS{\phi}(\theta)$, then either $\eta \in \newMS{\phi_1}(\theta)$ or
$\eta \in \newMS{\phi_2}(\theta)$, so by induction hypothesis either
${\cal I} \models_{\eta}\phi_1$ or ${\cal I} \models_{\eta}\phi_2$.
In both cases ${\cal I} \models_{\eta}\phi_1 \Or \phi_2$ holds.

If $error \not\in \newMS{\phi}(\theta)$, 
then for some ${\cal J}$-substitutions $\eta_1, \LL, \eta_k$

\[
\newMS{\phi_1}(\theta) = \C{\theta \eta_1, \LL, \theta \eta_k},
\]
where $k \geq 0$, for some ${\cal J}$-substitutions $\eta_{k+1}, \LL, \eta_{k+ \ell}$, 
\[
\newMS{\phi_2}(\theta) = \C{\theta \eta_{k+1}, \LL, \theta \eta_{k+ \ell}},
\]
where $\ell \geq 0$, and 
\[
\newMS{\phi_1 \Or \phi_2}(\theta)  = \C{\theta \eta_1, \LL, \theta \eta_{k+ \ell}}.
\]
By induction hypothesis both
\[
{\cal I} \models \phi_1 \theta \lra 
\bigvee_{i = 1}^{k}
\te {\bf y}_i \newhat{\eta_i}
\]
and
\[
{\cal I} \models \phi_2 \theta \lra 
\bigvee_{i = k+1}^{k + \ell}
\te {\bf y}_i \newhat{\eta_i}
\]
for appropriate  sequences of variables ${\bf y}_i$. So
\[
{\cal I} \models (\phi_1 \Or \phi_2) \theta \lra 
\bigvee_{i = 1}^{k + \ell}
\te {\bf y}_i \newhat{\eta_i}.
\]
\III

\NI
$\phi$ is $\neg \phi_1$.

If $\eta \in \newMS{\phi}(\theta)$, then 
$\eta = \theta$ and $\newMS{\phi_1}(\theta) = \ES$.
By induction hypothesis ${\cal I} \models_{\theta} \neg \phi_1$,
i.e., ${\cal I} \models_{\eta} \neg \phi_1$.

If $error \not\in \newMS{\phi}(\theta)$, then either
$\newMS{\phi}(\theta) = \C{\theta}$ or $\newMS{\phi}(\theta) = \ES$.
In the former case $\newMS{\phi}(\theta) = \C{\theta \varepsilon}$, so
$\newMS{\phi_1}(\theta) = \ES$. By induction hypothesis ${\cal I}
\models_{\theta} \neg \phi_1$, i.e., ${\cal I} \models (\neg
\phi_1)\theta \lra \newhat{\varepsilon}$, since $\newhat{\varepsilon}$
is vacuously true.  In the latter case $\theta \in
\newMS{\phi_1}(\theta)$, so by induction hypothesis ${\cal I}
\models_{\theta} \phi_1$, i.e., ${\cal I} \models (\neg \phi_1)\theta
\lra falsum$.
\III

\NI
$\phi$ is $\te x \: \phi_1$.

If $\eta \in \newMS{\phi}(\theta)$, then $\eta \in
DROP_y(\newMS{\phi_1 \C{x/y}}(\theta))$ for some fresh variable $y$.
So either (if $y$ is not in the domain of $\eta$) $\eta \in
\newMS{\phi_1 \C{x/y}}(\theta)$ or for some ${\cal J}$-term $s$ we
have $\eta \uplus \C{y/s} \in \newMS{\phi_1 \C{x/y}}(\theta)$.  By
induction hypothesis in the former case ${\cal I} \models_{\eta}
\phi_1 \C{x/y}$ and in the latter case ${\cal I} \models_{\eta \uplus
\{y/s\}} \phi_1 \C{x/y}$.  In both cases 
${\cal I} \models  \te y \: (\phi_1  \C{x/y} \eta)$, so,
since $y$ is fresh, 
${\cal I} \models  (\te y \: \phi_1  \C{x/y}) \eta$
and consequently
${\cal I} \models  (\te x \: \phi_1 ) \eta$, i.e., 
${\cal I} \models_{\eta}  \te x \: \phi_1 $.

If $error \not\in \newMS{\phi}(\theta)$, then 
$error \not\in \newMS{\phi_1 \C{x/y}}(\theta)$, as well,
where $y$ is a fresh variable.
By induction hypothesis 
\begin{equation}
{\cal I} \models \phi_1 \C{x/y} \theta \lra \bigvee_{i = 1}^{k} \te {\bf y}_{i} \newhat{\eta_i},
\label{equ:exists}
\end{equation}
where 
\begin{equation}
\newMS{\phi_1 \C{x/y}}(\theta) = \C{\theta \eta_1, \LL, \theta \eta_k}
\label{equ:sem}
\end{equation}
and for $i \in [1..k]$ $\ {\bf y}_i$ is a sequence of variables that
appear in the range of $\eta_i$.

Since  $y$ is fresh, we have
${\cal I} \models \te y \: (\phi_1 \C{x/y} \theta) \lra 
(\te y \: \phi_1 \C{x/y}) \theta$ and
${\cal I} \models (\te y \: \phi_1 \C{x/y}) \theta \lra 
(\te x \: \phi_1) \theta$. So (\ref{equ:exists}) implies
\[
{\cal I} \models (\te x \: \phi_1) \theta \lra 
\bigvee_{i = 1}^{k} \te y \te {\bf y}_{i} \newhat{\eta_i}.
\]

But for $i \in [1..k]$
\[
{\cal I} \models \te y \newhat{\eta_i} \lra \te y \newhat{DROP_{y}(\eta_i)},
\]
since if $y/s \in \eta_i$, then the variable $y$ does not appear in $s$.
So
\begin{equation}
{\cal I} \models (\te x \: \phi_1) \theta \lra 
\bigvee_{i = 1}^{k} \te {\bf y}_{i} \te y \newhat{DROP_{y}(\eta_i)}.
\label{equ:equiv}
\end{equation}
Now, by (\ref{equ:sem})
\[
\newMS{\te x \: \phi_1}(\theta) = \C{DROP_{y}(\theta \eta_1), \LL, DROP_{y}(\theta \eta_k)}.
\]
But $y$ does not occur in $\theta$, so we have for $i \in [1..k]$
\[
DROP_{y}(\theta \eta_i) = \theta DROP_{y}(\eta_i)
\]
and consequently
\[
\newMS{\te x \: \phi_1}(\theta) = \C{\theta DROP_{y}(\eta_1), \LL, \theta DROP_{y}(\eta_k)}.
\]
This by virtue of (\ref{equ:equiv}) concludes the proof.
\HB
\VV

Informally, $(i)$ states that every computed answer substitution of
$\phi \theta$ validates it.  It is useful to point out that $(ii)$ is a
counterpart of Theorem 3 in Clark \cite{Cla78}. Intuitively, it states
that a query is equivalent to the disjunction of its computed answer
substitutions written out in an equational form (using the
$\newhat{\eta}$ notation). In our case this property holds only if
$error$ is not a possible outcome. Indeed, if $\newMS{s=t}(\theta) =
\C{error}$, then nothing can be stated about the status of the
statement ${\cal I} \models (s =t) \theta$.

Note that in case $error \not\in \newMS{\phi}(\theta)$, $(ii)$ implies
$(i)$ by virtue of Note \ref{not:equality}.  On the other hand, if
$error \in \newMS{\phi}(\theta)$, then $(i)$ can still be applicable
while $(ii)$ not.

Additionally existential quantifiers have to be used in an appropriate
way. The formulas of the form $\te {\bf y} \newhat{\eta}$ also appear
in Maher \cite{Mah88} in connection with a study of the decision
procedures for the algebras of trees.  In fact, there are some
interesting connections between this paper and ours that could be
investigated in a closer detail.

\section{Conclusions and Future Work}
\label{sec:conclusions}

In this paper we provided a denotational semantics to first-order
logic formulas.  This semantics is a counterpart of the operational
semantics introduced in Apt and Bezem \cite{AB99}. The important
difference is that we provide here a more general treatment of equality
according to which a non-ground term can be assigned to a variable.
This realizes logical variables in the framework of Apt and Bezem
\cite{AB99}.  This feature led to a number of complications in the
proof of the Soundness Theorem \ref{thm:soundness}.

One of the advantages of this theorem is that it allows us to reason
about the considered program simply by comparing it to the formula
representing its specification.  In the case of operational semantics
this was exemplified in Apt and Bezem \cite{AB99} by showing how to
verify non-trivial \almazero{} programs that do not include
destructive assignment.

Note that it is straightforward to extend the semantics here provided
to other well-known programming constructs, such as destructive
assignment, {\bf while} construct and recursion.  However, as soon as
a destructive assignment is introduced, the relation with the
definition of truth in the sense of Soundness Theorem
\ref{thm:soundness} is lost and the just mentioned approach to program
verification cannot be anymore applied.  In fact, the right approach
to the verification of the resulting programs is an appropriately designed
Hoare's logic or the weakest precondition semantics.

The work here reported can be extended in several directions.  First
of all, it would be useful to prove equivalence between the
operational and denotational semantics.  Also, it would interesting to
specialize the introduced semantics to specific interpretations for
which the semantics could generate less often an error.  Examples are
Herbrand interpretations for an arbitrary first-order language in
which the meaning of equalities could be rendered using most general
unifiers, and the standard interpretation over reals for the language
defining linear equations; these equations can be handled by means of
the usual elimination procedure.  In both cases the equality could be
dealt with without introducing the {\em error\/} state at all. 

Other possible research directions were already mentioned in Apt and
Bezem \cite{AB99}. These involved addition of recursive procedures, of
constraints, and provision of a support for automated verification of
programs written in \almazero{}.  The last item there mentioned,
relation to dynamic predicate logic, was in the meantime extensively
studied in the work of van Eijck \cite{vE98} who, starting with Apt
and Bezem \cite{AB99}, defined a number of semantics for dynamic
predicate logic in which the existential quantifier has a different,
dynamic scope.  This work was motivated by applications in natural
language processing.

\section*{Acknowledgments}
Many thanks to Marc Bezem for helpful discussions on the subject of
this paper.

\bibliographystyle{alpha}

\newcommand{\etalchar}[1]{$^{#1}$}


\end{document}